\documentclass[preprint,tightenlines,aps,showpacs]{revtex4}
\usepackage{amsmath}
\usepackage{amssymb}
\usepackage{epsfig}

\begin{document}

\title{Examination of the H dibaryon within a chiral constituent quark model}
\author{T.F.~Caram\'es and A.~Valcarce}
\affiliation{Departamento de F\'\i sica Fundamental, Universidad de Salamanca, E--37008
Salamanca, Spain}
\date{\today}

\begin{abstract}
We perform a coupled--channel calculation of the H dibaryon within a chiral constituent quark model. The problem is solved within a quark model constrained by the experimental data of strangeness --1 and --2 two--baryon systems. We examine in detail the role played by the different contributions of the interacting potential as well as the number of coupled channels considered. Special attention has been payed to the parameter dependence, flavor symmetry breaking and spatial configurations. The value extracted for the binding energy of the H dibaryon, being compatible with the restrictions imposed by the Nagara event, falls within a plausible extrapolation of recent lattice QCD results.
\end{abstract}

\pacs{12.39.Jh,12.39.Pn,14.20.Jn,14.20.Pt}
\maketitle

\section{Introduction}
Very recently the H dibaryon was put back on the agenda by lattice QCD calculations of different collaborations, NPLQCD~\cite{Beane:2010hg} and HAL QCD~\cite{Inoue:2010es}, providing evidence for a bound state at pion masses larger than the physical ones. It was first proposed thirty years ago~\cite{Jaffe:1976yi} as a spin and flavor singlet composed of six quarks ($uuddss$). Such a proposal emerged on the context of the MIT bag model~\cite{Chodos:1974je}. When applied to all six--quark systems it predicted the existence of only one stable dihyperon with $J^P = 0^+$ and a mass of 2150 MeV, 81 MeV below the $\Lambda\Lambda$ threshold, therefore strongly bound. Moreover, being the lightest particle of a two-baryon system with strangeness ($\hat{S}$) --2, it would be stable against the strong interaction and would necessarily decay weakly. It was assumed to be a single hadron made of six quarks squeezed in a small region and not a two--baryon state bound in an $S$ wave like the deuteron. If this were the case it would open the door to the exotic states, i.e., hadrons that do not fit in the standard $\bar{q}q$ or $qqq$ configurations. Otherwise, the fact that the H dibaryon is the most promising candidate to be the second dibaryonic state after the deuteron has generated a lot of expectation.

On the experimental side, there are very few data in the $\hat{S}=-2$ sector, coming from the inelastic $\Xi^- p \to \Lambda\Lambda$ cross section at a lab momentum of around 500 MeV/c, and from the elastic $\Xi^- p \to \Xi^- p$ and inelastic $\Xi^- p \to \Xi^0 n$ cross sections for lab momenta in the range of 500 -- 600 MeV/c~\cite{Ahn:2005jz,Tamagawa:2001tk,Yamamoto:2001tn}. Thus, the relevant information we have is indirect and comes from double--$\Lambda$ hypernuclei. Their binding energies, $B_{\Lambda\Lambda}$, provide upper limits for that of the H dibaryon, i.e., $B_H < B_{\Lambda\Lambda}$. The first hypernuclear events are quite old and admit several interpretations~\cite{Danysz:1963zz,Prowse:1966nz,Aoki:1991ip}. In 2001 it was reported the so--called Nagara event~\cite{Takahashi:2001nm}, interpreted uniquely as the sequential decay of ${}^6_{\Lambda\Lambda}$He emitted from a $\Xi^-$--hyperon nuclear capture at rest. The mass and the values of $B_{\Lambda\Lambda}$ and of the $\Lambda\Lambda$ interaction energy, $\Delta B_{\Lambda\Lambda}$, were determined without ambiguities. The small value of $\Delta B_{\Lambda\Lambda}$ suggested an attraction weaker than the one previously estimated. It also gave the most stringent constraint to the mass of the H dibaryon to date, i.e., $M_H > 2223.7$ MeV at a 90\% confidence level. It took almost one decade, but four more double--$\Lambda$ hypernuclear events were reported, from KEK E176 and E373 experiments~\cite{Nakazawa:2010zza}, still with preliminary results. All the details are summarized in Table~\ref{hip}. 
\begin{table}[b]\caption{Reported double hypernuclear events.}
\begin{center}
\begin{tabular}{c  c c c}
\hline\hline
Event & Nuclide & $B_{\Lambda\Lambda}$ (MeV) & $\Delta B_{\Lambda\Lambda}$ (MeV) \\
\hline
1963          & ${}_{\Lambda\Lambda}^{10}$Be & 17.7 $\pm$ 0.4  & 4.3  $\pm$ 0.4   \\
1966          & ${}_{\Lambda\Lambda}^6$He    & 10.9 $\pm$ 0.5  & 4.7  $\pm$ 1.0   \\
1991          & ${}_{\Lambda\Lambda}^{13}$B  & 27.5 $\pm$ 0.7  & 4.8  $\pm$ 0.7   \\
NAGARA        & ${}_{\Lambda\Lambda}^6$He    & 7.13 $\pm$ 0.87 & 1.0  $\pm$ 0.2   \\
MIKAGE        & ${}_{\Lambda\Lambda}^6$He    & 10.06 $\pm$ 1.72& 3.82 $\pm$ 1.72  \\
DEMACHIYANAGI & ${}_{\Lambda\Lambda}^{10}$Be & 11.90 $\pm$ 0.13& --1.52 $\pm$ 0.15 \\
HIDA          & ${}_{\Lambda\Lambda}^{11}$Be & 20.49 $\pm$ 1.15&  2.27 $\pm$ 1.23 \\
              & ${}_{\Lambda\Lambda}^{12}$Be & 22.23 $\pm$ 1.15&   --             \\
E176          & ${}_{\Lambda\Lambda}^{13}$Be &  23.3 $\pm$ 0.7 & 0.6 $\pm$ 0.8 \\
\hline\hline
\end{tabular}
\end{center}\label{hip}
\end{table}
The future E07 experiment from J--PARC~\cite{Nakazawa:2010zza} is expected to improve our knowledge on the $\hat{S}=$--2 sector, giving ten times more events. 

From the theoretical point of view, many approaches have been performed, their predictions for the binding of the H spreading over a wide range of energies~\cite{Sakai:1999qm}. Recent lattice results produced by the NPLQCD~\cite{Beane:2010hg} and HAL QCD~\cite{Inoue:2010es} Collaborations  found a bound H dibaryon for non--physical values of the pion mass ($m_{\pi}$ = 837 MeV and $m_\pi$ = 670 $\to$ 1010 MeV, respectively). When performing quadratic and linear extrapolations to the physical point~\cite{Beane:2011xf}, a bound dibaryon (around 7 MeV) and a H at threshold, respectively, are predicted. Also presented in Ref.~\cite{Beane:2011xf} are preliminary results for $m_{\pi}$ = 230 MeV, much closer to the physical pion mass, pointing to a H dibaryon at threshold, as also experimentally suggested by the enhancement of the $\Lambda\Lambda$ production near threshold found in Ref.~\cite{Yoon:2007aq}. 

In this work we present the first approach to the H dibaryon within a constituent quark model constrained by the experimental data of the $\hat{S}=$ --1 and --2 cross sections~\cite{Garcilazo:2007ss,Valcarce:2010kz}. 
To make our results more robust and significative, we scrutinize all channels in the $\hat{S}=$ --2 sector, paying due attention to the H dibaryon, $(T,S)=(0,0)$.
The paper is organized as follows. In Section II we provide a brief description of the constituent quark model and the formalism to study the coupled--channel problem. In Section III we present and analyze our results. Finally, we summarize our conclusions in Section IV.

\section{The chiral constituent quark model}
\label{secII} 
The baryon--baryon interactions needed for the study of the H dibaryon are computed from a chiral constituent quark model (CCQM)~\cite{Valcarce:2005em}. Baryons are described as clusters of three effective constituent quarks, their mass coming from the spontaneous chiral symmetry breaking. The CCQM was first applied to the study of the nonstrange SU(2)$\times$SU(2) sector, describing the baryon spectroscopy and the NN interaction in a consistent manner~\cite{Valcarce:2005em}. Such a success was due to the choice of the adequate mechanisms in the description of the quark--quark--meson interaction: quark antisymmetry plus a perturbative short--range force together with a microscopic non--perturbative chiral interaction at medium and long distances. They allow a unified treatment of the one--, two-- and three--body systems, with a reduced and unique set of parameters. Later on, a generalization to SU(3)$\times$SU(3) was done to perform a systematic and detailed analysis of the $q \bar{q}$ spectrum~\cite{Vijande:2004he}. The interaction between quarks was given by (Model I from now on):
\begin{equation}
 V_{qq} (\vec r) = V_{\rm CON}(\vec r) + V_{\rm OGE}(\vec r) + V_{\sigma}(\vec r) + V_{\rm PSE}(\vec r)\,,
\label{potenqq}
\end{equation}
where $V_{\rm CON}$ is a confinement term that represents the nonperturbative aspects of QCD, $V_{\rm OGE}$ is the one--gluon exchange (OGE) potential, obtained through the nonrelativistic reduction of the quark--quark--gluon interaction diagram in QCD, $V_{\sigma}$ is the one--sigma--exchange potential and $V_{\rm PSE}$ stands for the chiral potential, associated to the exchange of pseudoscalar Goldstone bosons. It comprises one--pion, one--eta and one--kaon exchanges. When studying the $\hat{S} = -1$ $\Lambda p$ cross section, it was noticed that a better approximation to the scalar interaction was needed, considering the nonet of scalar mesons,
that comprises a singlet (denoted by $\sigma_0$) and an octet that will be denoted by $V_{\rm SCE}$, giving thus rise to Model II: 
\begin{equation}
 V_{qq} (\vec r) = V_{\rm CON}(\vec r) + V_{\rm OGE}(\vec r) + V_{\sigma_0}(\vec r) + V_{\rm SCE}(\vec r) + V_{\rm PSE}(\vec r)\,,
\label{potqq}
\end{equation}
where the relation between $\sigma$ and $\sigma_0$ is given by $\sigma = \cos \theta_s \, \sigma_0 + \sin \theta_s \, \sigma_8$. The present model has been used in Ref.~\cite{Garcilazo:2007ss} to study two-- and three--baryon systems with strangeness --1 giving a nice description of the hypertriton. It has also been used to study the strangeness --2 two--body scattering cross sections~\cite{Valcarce:2010kz}. A reasonable fit to the experimental data of the elastic $\Xi^- p$ and the inelastic $\Xi^- p \to \Xi^0 n$ and $\Xi^- p \to \Lambda\Lambda$ cross sections reported in Refs.~\cite{Ahn:2005jz,Tamagawa:2001tk,Yamamoto:2001tn} was obtained. 

In order to derive the local $B_1B_2\to B_3B_4$ potentials from the basic $qq$ interaction we use a Born--Oppenheimer approximation. Explicitly, the potential is calculated as follows,
\begin{equation}
V_{B_1B_2 (L \, S \, T) \rightarrow B_3B_4
(L^{\prime}\, S^{\prime}\, T)} (R) =
\xi_{L \,S \, T}^{L^{\prime}\, S^{\prime}\, T}
(R) \, - \, \xi_{L \,S \,
T}^{L^{\prime}\, S^{\prime}\, T} (\infty) \, , 
\label{Poten1}
\end{equation}
\noindent where
\begin{equation}
\xi_{L \, S \, T}^{L^{\prime}\, S^{\prime}\, T}
(R) \, = \, {\frac{{\left
\langle \Psi_{B_3B_4 }^{L^{\prime}\,
S^{\prime}\, T} ({\vec R}) \mid
\sum_{i<j=1}^{6} V_{q_iq_j}({\vec r}_{ij}) \mid
\Psi_{B_1B_2 }^{L \, S \, T} ({\vec R%
}) \right \rangle} }{{\sqrt{\left \langle
\Psi_{B_3B_4 }^{L^{\prime}\,
S^{\prime}\, T} ({\vec R}) \mid \Psi_{B_3B_4
}^{L^{\prime}\, S^{\prime}\, T} ({%
\vec R}) \right \rangle} \sqrt{\left \langle
\Psi_{B_1B_2}^{L \, S \, T} ({\vec %
R}) \mid \Psi_{B_1B_2}^{L \, S \, T} ({\vec R})
\right \rangle}}}} \, .
\label{Poten2}
\end{equation}
In the last expression the quark coordinates are integrated out keeping $R$ fixed, the resulting interaction being a function of the $B_i-B_j$  relative distance. The wave function $\Psi_{B_iB_j}^{L \, S \, T}({\vec R})$ for the two--baryon system is discussed in detail in Ref.~\cite{Valcarce:2005em}. This formalism allows us to isolate different contributions and/or diagrams, making it easy to analyze the results.

Once we have the two--baryon interactions, we switch to solve the two--body coupled--channel problem. Let us start from a physical system made of two baryons, $B_1$ and $B_2$ ($B_i=\Lambda, N, \Xi, \Sigma$), with isospin, spin and parity quantum numbers $(I)J^{P}$ in a relative $S$ state. They interact through a potential $V$ that contains a tensor force. Then, in general, there is a coupling to the $B_1 B_2$ $D-$wave and to any other two baryon system ($\Lambda \Lambda$, $N \Xi$, $\Sigma\Sigma$) that can couple to the same quantum numbers $(I)J^{P}$. Thus, if we denote $\Lambda\Lambda \equiv D_1 $, $N \Xi \equiv D_2$ and $\Sigma\Sigma \equiv D_3$, the Lippmann--Schwinger equation for the $\hat{S} = -2$ $B_1B_2$ scattering becomes 
\begin{eqnarray}
t_{\alpha\beta;ji}^{\ell_\alpha s_\alpha,
\ell_\beta s_\beta}(p_\alpha,p_\beta;E)& = & 
V_{\alpha\beta;ji}^{\ell_\alpha s_\alpha,
\ell_\beta s_\beta}(p_\alpha,p_\beta)+
\sum_{\begin{array}{c}{\scriptstyle \gamma=D_k}
\\{\scriptstyle
(k=1,2,3)}\end{array}}\sum_{\ell_\gamma=0,2} 
\int_0^\infty p_\gamma^2 dp_\gamma
V_{\alpha\gamma;ji}^{\ell_\alpha s_\alpha,
\ell_\gamma s_\gamma}
(p_\alpha,p_\gamma) \nonumber \\
& \times& \, G_\gamma(E;p_\gamma)
t_{\gamma\beta;ji}^{\ell_\gamma s_\gamma,
\ell_\beta s_\beta}
(p_\gamma,p_\beta;E) \,\,\,\, , \,
\alpha,\beta=D_1,D_2,D_3 \,\, ,
\label{eq0}
\end{eqnarray}
where $t$ is the two--body scattering amplitude, $j$, $i$, and $E$ are the angular momentum, isospin and energy of the system, respectively; $\ell_{\alpha} s_{\alpha}$, $\ell_{\gamma} s_{\gamma}$, and $\ell_{\beta} s_{\beta }$ are the initial, intermediate, and final orbital angular momentum and spin, respectively,  and $p_\gamma$ is the relative momentum of the two--body system $\gamma$. The propagators $G_\gamma(E;p_\gamma)$ are given by 
\begin{equation}
G_\gamma(E;p_\gamma)=\frac{2
\mu_\gamma}{k^2_\gamma-p^2_\gamma + i \epsilon}
\, ,
\end{equation}
with
\begin{equation}
E=\frac{k^2_\gamma}{2 \mu_\gamma} \, ,
\end{equation}
where $\mu_\gamma$ is the reduced mass of the two--body system $\gamma$. For bound--state problems $E < 0$ so that the singularity of the propagator is never touched and we can forget the $i\epsilon$ in the denominator. If we make the change of variables 
\begin{equation}
p_\gamma = b \frac{1+x_\gamma}{1-x_\gamma},
\label{eq2a}
\end{equation}
where $b$ is a scale parameter, and the same for $p_\alpha$ and $p_\beta$, we can write Eq.~(\ref{eq0}) as 
\begin{eqnarray}
t_{\alpha\beta;ji}^{\ell_\alpha s_\alpha, \ell_\beta s_\beta}(x_\alpha,x_\beta;E)& = & 
V_{\alpha\beta;ji}^{\ell_\alpha s_\alpha, \ell_\beta s_\beta}(x_\alpha,x_\beta)+ 
\sum_{\begin{array}{c}{\scriptstyle \gamma=D_k} 
\\{\scriptstyle
(k=1,2,3)}\end{array}}\sum_{\ell_\gamma=0,2} 
\int_{-1}^1 b^2\left(1+x_\gamma \over 1-x_\gamma
\right)^2 \,\, {2b \over (1-x_\gamma)^2}
dx_\gamma \nonumber \\
&\times & V_{\alpha\gamma;ji}^{\ell_\alpha
s_\alpha, \ell_\gamma s_\gamma}
(x_\alpha,x_\gamma) \, G_\gamma(E;p_\gamma) \,
t_{\gamma\beta;ji}^{\ell_\gamma s_\gamma,
\ell_\beta s_\beta}
(x_\gamma,x_\beta;E) \, .
\label{eq3}
\end{eqnarray}
We solve this equation by replacing the integral from $-1$ to $1$ by a Gauss--Legendre quadrature which results in the set of linear equations 
\begin{equation}
\sum_{\begin{array}{c}{\scriptstyle \gamma=D_k}
\\{\scriptstyle (k=1,2,3)}\end{array}}
\sum_{\ell_\gamma=0,2}\sum_{m=1}^N
M_{\alpha\gamma;ji}^{n \ell_\alpha s_\alpha, m
\ell_\gamma s_\gamma}(E) \, 
t_{\gamma\beta;ji}^{\ell_\gamma s_\gamma,
\ell_\beta s_\beta}(x_m,x_k;E) =  
V_{\alpha\beta;ji}^{\ell_\alpha s_\alpha,
\ell_\beta s_\beta}(x_n,x_k) \, ,
\label{eq4}
\end{equation}
with
\begin{eqnarray}
M_{\alpha\gamma;ji}^{n \ell_\alpha s_\alpha, m
\ell_\gamma s_\gamma}(E)
& = & \delta_{nm}\delta_{\ell_\alpha
\ell_\gamma} \delta_{s_\alpha s_\gamma}
- w_m b^2\left(1+x_m \over 1-x_m\right)^2{2b
\over (1-x_m)^2} \nonumber \\
& \times & V_{\alpha\gamma;ji}^{\ell_\alpha
s_\alpha, \ell_\gamma s_\gamma}(x_n,x_m) 
\, G_\gamma(E;{p_\gamma}_m),
\label{eq5}
\end{eqnarray}
  and where $w_m$ and $x_m$ are the weights and abscissas of the Gauss--Legendre quadrature, while ${p_\gamma}_m$ is obtained by putting $x_\gamma=x_m$ in Eq.~(\ref{eq2a}). If a bound state exists at an energy $E_B$, the determinant of the matrix $M_{\alpha\gamma;ji}^{n \ell_\alpha s_\alpha, m \ell_\gamma s_\gamma}(E_B)$ vanishes, i.e., $\left|M_{\alpha\gamma;ji}(E_B)\right|=0$.
We took the scale parameter $b$  of Eq.~(\ref{eq2a}) as $b$ = 3 fm$^{-1}$ and used a Gauss--Legendre quadrature with $N=$ 20 points.
\begin{table}[b]\caption{Single channel binding energy (in MeV).}
\begin{center}
\begin{tabular}{c|c  c c}
\hline\hline
&$\quad B_{\Lambda\Lambda}\quad$&$\quad B_{N\Xi}\quad$&$\quad B_{\Sigma\Sigma}\quad$ \\
 \hline 
 Model I  & 1.9  &  32.7 & 24.2  \\
 Model II & --   &  0.1  &  0.1  \\
 \hline\hline
\end{tabular}
\end{center}\label{onechfred}
\end{table}
\section{Results}
\label{secIII}
The flavor singlet $(T,S) = (0,0)$ state in the $\hat{S} = -2$ sector comprises three coupled two--baryon channels, presenting rich coupling effects. If SU(3) were exact, one could write
\begin{equation}
\big|H \big> = \sqrt{\dfrac{1}{8}} |\Lambda\Lambda \rangle + \sqrt{\dfrac{4}{8}} |N \Xi \rangle - \sqrt{\dfrac{3}{8}} |\Sigma\Sigma \rangle \,.
\label{hbcon}
\end{equation} 
The presence of three transition potentials, namely $\Lambda\Lambda - N\Xi$, $\Lambda\Lambda - \Sigma\Sigma$ and $N \Xi - \Sigma\Sigma$, makes it necessary, to obtain the bound states of the system, to perform a coupled--channel calculation where six different interactions contribute: three diagonal potentials ($\Lambda\Lambda$, $N \Xi$ and $\Sigma\Sigma$) plus the three transitions just mentioned.

\begin{figure}[tbp]
\begin{center}
\vspace*{-3.3cm}
 \mbox{\psfig{figure=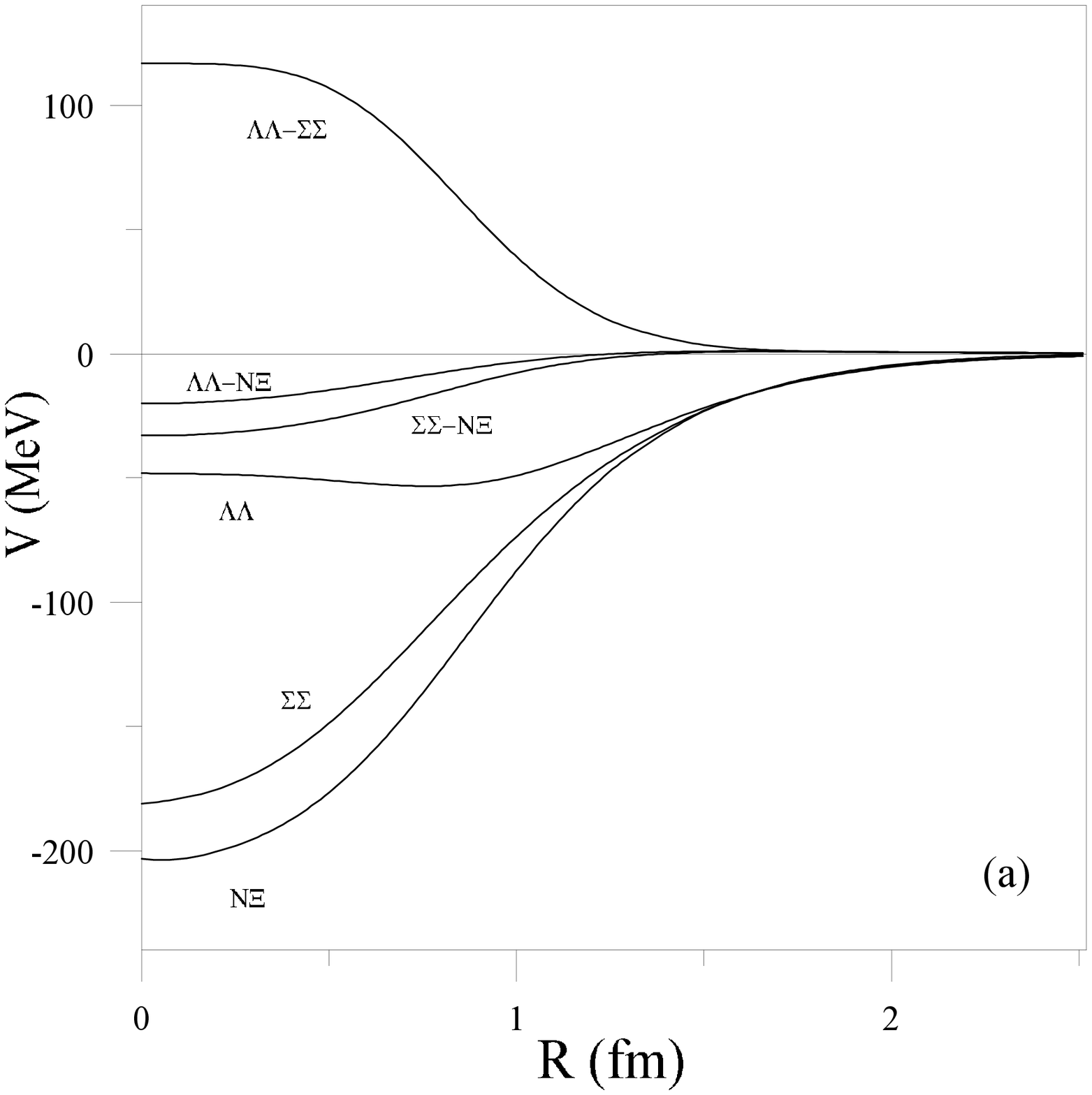,height=4.1in,width=2.7in}}
\hspace*{0.5cm}
 \mbox{\psfig{figure=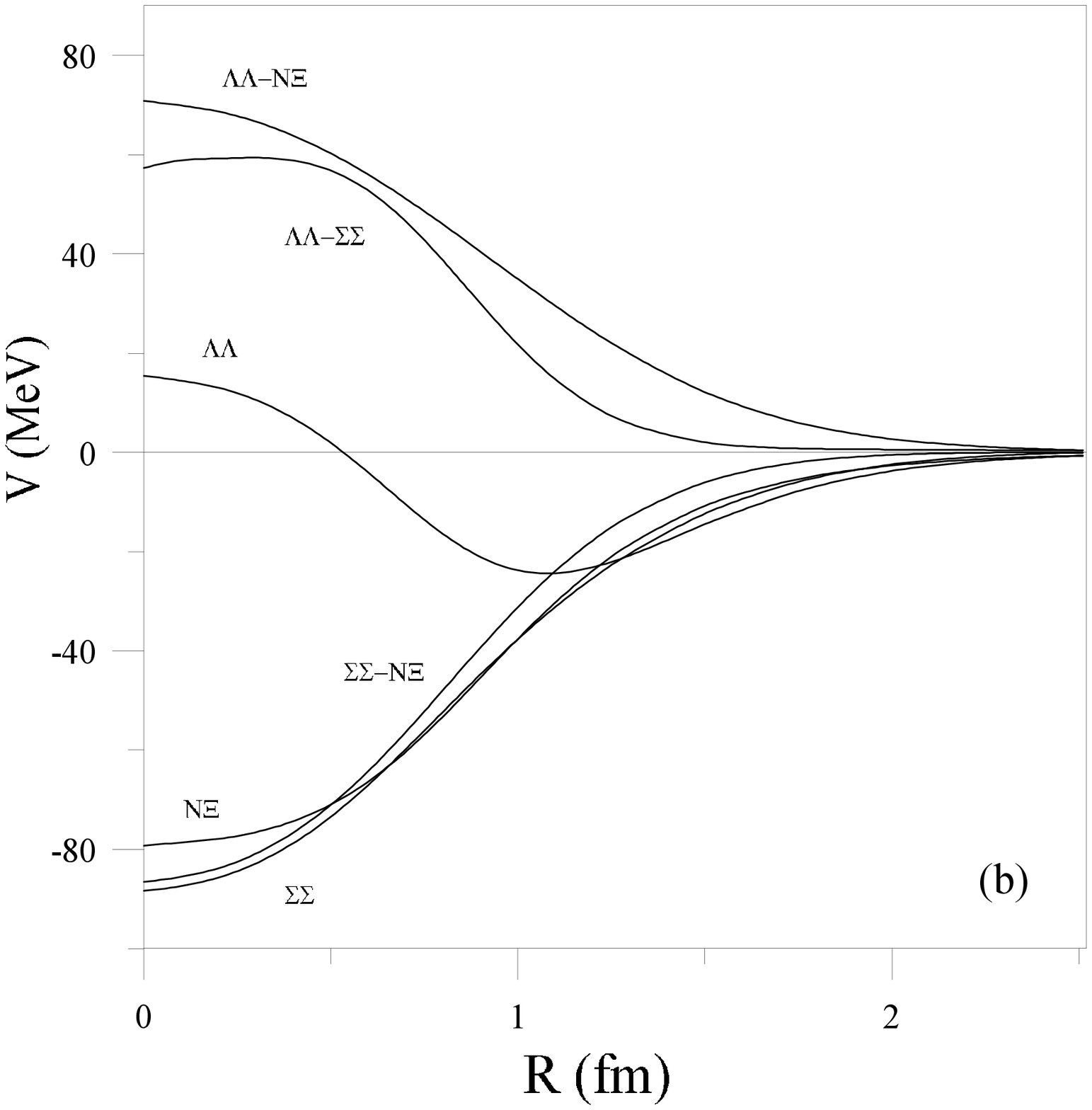,height=4.1in,width=2.7in}}
\vspace*{-0.8cm}
\caption{Interacting potentials contributing to the H dibaryon in ($\rm a$) Model I and ($\rm b$) Model II. \label{Hchannel}}
\end{center}
\end{figure}

Regarding the baryon--baryon interactions, let us only point out here a couple of remarkable features of interest for the H dibaryon study. First of all, the three diagonal potentials are attractive. The smaller attraction corresponds to the $\Lambda\Lambda$ potential, whereas $N \Xi$ and $\Sigma\Sigma$ show broader and deeper attractions, respectively. In order to know about the magnitude of these diagonal interactions, the Fredholm determinant --that tells the binding energy of a system-- can help. The binding energies of these three channels with respect to their own thresholds (still without any coupling), computed in Models I and II, are shown in Table~\ref{onechfred}. The six potentials contributing to the H dibaryon in Models I and II are displayed in Figs.~\ref{Hchannel}($\rm a$) and ($\rm b$). From them we shall proceed to solve the coupled--channel problem by enlarging the Hilbert space progressively, from $\{\Lambda\Lambda\}$ to $\{ \Lambda\Lambda, N \Xi\}$ and finally to $\{\Lambda\Lambda, N \Xi, \Sigma\Sigma\}$. The thresholds of these three channels are respectively 2231 MeV, 2257 MeV and 2381 MeV. 

In this way we are able to separate and to identify the modifications introduced by every baryon--baryon channel. The results, computed in Models I and II, are shown in Table~\ref{hilbert}.
\begin{table}[b]\caption{$\hat{S}=-2$, $(T,S)=(0,0)$ binding energy (in MeV) in the one--, two-- and three--channel approximations.}
\begin{center}
\begin{tabular}{c|c  c c}
\hline\hline
 & $\quad B_{\{\Lambda\Lambda\}} \quad$ & $\quad B_{\{\Lambda\Lambda, N\Xi\}} \quad$ & $ \quad B_{\{\Lambda\Lambda, N\Xi, \Sigma\Sigma \}} \quad$ \\
 \hline 
 Model I  &  1.9  &   8.7  &  10.0  \\
 Model II &  --   &   1.6  &  7.0  \\
 \hline\hline
\end{tabular}
\end{center}\label{hilbert}
\end{table}
When approximating the interaction by the $\Lambda\Lambda$ potential, one finds the H dibaryon slightly above threshold in Model II. As expected, due to the attractive character of the remaining channels, the coupling increases the binding energy of the dihyperon by several MeV. The low sensitivity of this binding when going from Model I to Model II contrasts with the important modification of the $\Lambda p$ cross section~\cite{Garcilazo:2007ss}. Hence a detailed analysis of the different contributions to the binding may be helpful. For that purpose we shall restrict ourselves to Model II, as it is the one that properly describes the $\hat{S}=-1$ and $-2$ experimental cross sections~\cite{Valcarce:2010kz}.

\subsection{Pieces of the interaction}
\noindent
A more precise analysis can be done by studying the different components of the interaction separately. Let us start with the chromomagnetic interaction consequence of the OGE, the basic mechanism originally proposed to bind the H~\cite{Jaffe:1976yi}. 
In the work by Oka {\it et al.}~\cite{Oka:1983ku}, using a one--gluon exchange plus a confining potential, a resonance 152 MeV above the $\Lambda\Lambda$ threshold was obtained. The large difference with respect to the original work~\cite{Jaffe:1976yi}, that gave $B_H =$ 81 MeV, was a consequence from the fact that Jaffe~\cite{Jaffe:1976yi} used the limit of flavor SU(3) symmetry and took for $\left(uuddss\right)$ the short--range correlation coefficients as for ordinary hadrons. Oka \textit{et al}.~\cite{Oka:1983ku} showed that in the dilute $\left(uuddss\right)$ system the strength of chromomagnetic effects is reduced as compared to ordinary hadrons.

\begin{figure}[tbp]
\begin{center}
\vspace*{-3.3cm}
 \includegraphics[scale=0.4]{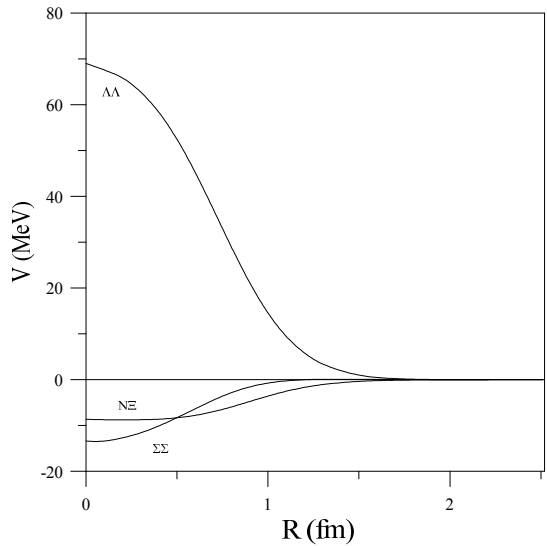}
 \vspace*{-1.3cm}
\caption{OGE interaction in the diagonal channels contributing to the H. \label{gluon}}
\end{center}
\end{figure}

Later on, several works employing one--gluon exchange potentials have approached the study of the H dibaryon offering again a wide range of results. Ref.~\cite{Shen:1999pf} took into account two possibilities: six--quark and two--cluster configurations. In the first one, even in the simplest case of OGE plus confinement, the H dibaryon was more than 300 MeV above threshold. In the second one, although the H was not yet bound, the energies of the resonance were much smaller. This was due to the repulsive character of the chromomagnetic  interaction in the $\Lambda\Lambda$ potential, not compensated by its attraction in the remaining channels. The very same feature was found by Nakamoto \textit{et al.}~\cite{Nakamoto:1997gh}, where, moreover, the repulsion in $\Lambda\Lambda$ was much stronger. 
On the other side, Stancu \textit{et al.}~\cite{Stancu:1997dq}, by performing a study of the short range part of the Goldstone boson exchange interaction, lacking thus the OGE potential, found a resonance more than 800 MeV above threshold.  
Finally, in a chiral quark model study with six--quark $(0s)^6$ as well as two--baryon configurations done by Shimizu {\it et al.}~\cite{Shimizu:1999fq}, three different cases were considered: a pure one--gluon exchange model supplemented by the long range Yukawa part of the pseudoscalar interaction, a hybrid model containing both OGE and chiral pseudoscalar plus scalar interactions, and a pure chiral model. Among them, the pure one--gluon exchange model produced by far the largest binding energy.
\begin{table}[t]\caption{Character of the strangeness --2 two--baryon interaction in the one--, two-- and three--channel approximations, for different quark--quark interactions. R indicates repulsion, WA weak attraction, $A(N)$ indicates attraction, being $N$ the binding energy in MeV. PSE stands for the pseudoscalar exchange, OGE for gluon, SCE for the scalar octet and $\sigma_0$ for the scalar singlet as indicated in Eq.~(\ref{potqq}).}
\begin{center}
\begin{tabular}{l|c  c c}
\hline\hline
 &  $\quad B_{\{\Lambda\Lambda\}} \quad$ & $\quad B_{\{\Lambda\Lambda, N\Xi\}} \quad$ & $ \quad B_{\{\Lambda\Lambda, N\Xi, \Sigma\Sigma \}} \quad$ \\
 \hline 
 OGE                           &   R       &    R      &    R      \\
 PSE                           &   R       &    R      &    R      \\
 $\sigma_0$                    &  A(5.4)   &   A(5.4)  &  A(8.3)   \\
 SCE                           &   WA      &   WA      &   R       \\
 OGE + PSE                     &   R       &   R       &    R      \\
 $\sigma_0$ + PSE              &  A(0.1)   &   A(0.1)  &  A(0.4)   \\ 
 OGE + $\sigma_0$              &  A(0.1)   &  A(0.1)   &  A(0.6)   \\
 OGE + PSE + $\sigma_0$        &   WA      &    WA     &   WA      \\
 OGE + PSE + $\sigma_0$ + SCE  &   WA      &   A(1.6)  &  A(7.0)   \\
 \hline\hline
\end{tabular}
\end{center}\label{contrib}
\end{table}

In our case the situation is qualitatively similar to Refs.~\cite{Nakamoto:1997gh}. The OGE interaction in the three diagonal channels is plotted in Fig.~\ref{gluon}. The repulsive character of the gluon exchange in $\Lambda\Lambda$ exceeds by far the attraction present in $\Sigma\Sigma$, and even more in $N \Xi$. Therefore if one looks at the binding energy with only gluon interaction, one finds that the Fredholm determinant is larger than 1, what is associated to a repulsive force. This is a general conclusion valid for any value of $\alpha_s$, since it is a multiplicative constant that appears in the three channels and the ratio between the repulsion in $\Lambda\Lambda$ and the attraction in $\Sigma\Sigma$ is preserved. Indeed, such an approximated ratio ($|V_{\rm OGE}^{\Lambda\Lambda} (R = 0)| / |V_{\rm OGE}^{\Sigma\Sigma} (R = 0)| \simeq 5$) is also found in~\cite{Nakamoto:1997gh}, where a SU(6) quark model is used, in spite of the huge difference in the quark--gluon coupling constant, taken there to be $\alpha_s = $ 1.52, almost one order of magnitude larger. The global effect of the chromomagnetic interaction is thus found to be repulsive and therefore one has to keep looking for the responsible of the binding.

In order to appreciate the overall features of any interaction, the $\hat{S}=-2$, $(T,S)=(0,0)$ two--baryon system has been solved in the one-- two-- and three--channel approaches by including only the interactions indicated in each line of Table~\ref{contrib}. This gives us information on the character of the interaction involved, whether it is repulsive, weakly attractive or attractive enough to bind the system. 
The pseudoscalar interaction generates repulsion, and it contributes only to the $\Lambda\Lambda$ channel, being negligible in the others. Therefore in a model containing only PSE, or OGE + PSE interactions there is not a bound state. The same situation occurs in Refs.~\cite{Straub:1988mz}, where a resonance 26 MeV above threshold is found in a OGE + PSE model. However, the addition of a scalar exchange at the baryonic level provides large attraction so that the binding becomes $B_H =$ 20 MeV. 
In Ref.~\cite{Stancu:1997dq} the pseudoscalars also introduce repulsion at short distances, but weaker than in the $NN$ interaction, in such a way that a bound H particle --provided there were medium range attraction as the $\sigma$ implies-- is not discarded. Repulsive pseudoscalar contribution is also found in~\cite{Shimizu:1999fq}, together with an attractive $\sigma$ piece, for a $(0s)^6$ configuration. When considering the two--cluster configuration with a pure chiral content  the bound state does not appear. The reason for that is not a repulsive character of the chiral meson exchange but the lack of strength. Because of that, an increment of the chiral coupling constant in order to properly describe the $^1S_0 \, NN$ phase shifts is successful for binding the system. 
\begin{figure}[tbp]
\begin{center}
\vspace*{-1.3cm}
\hspace*{-0.4cm}\mbox{\psfig{figure=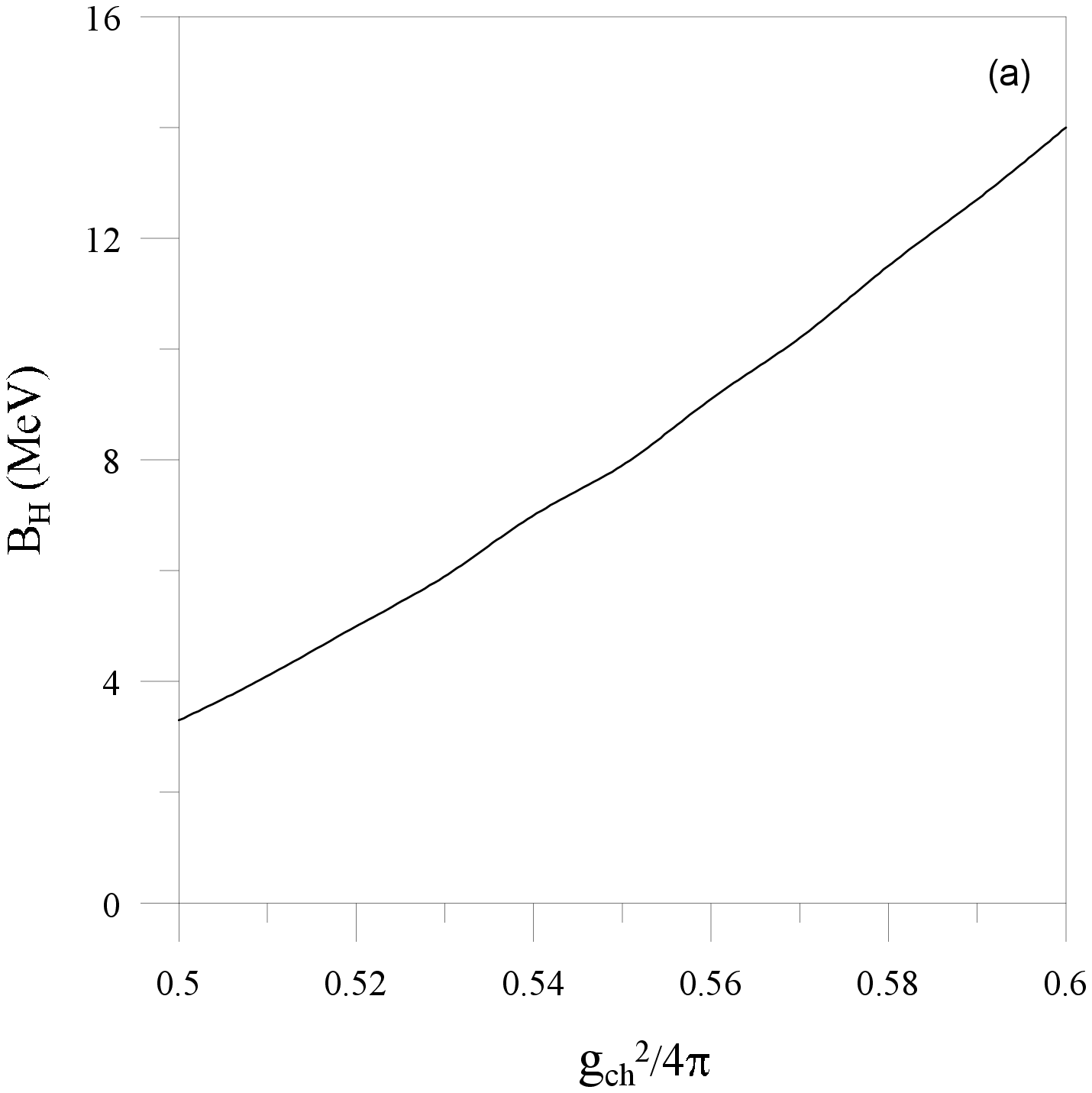,height=4.5in,width=3.1in}}
\hspace*{0.4cm}
 \mbox{\psfig{figure=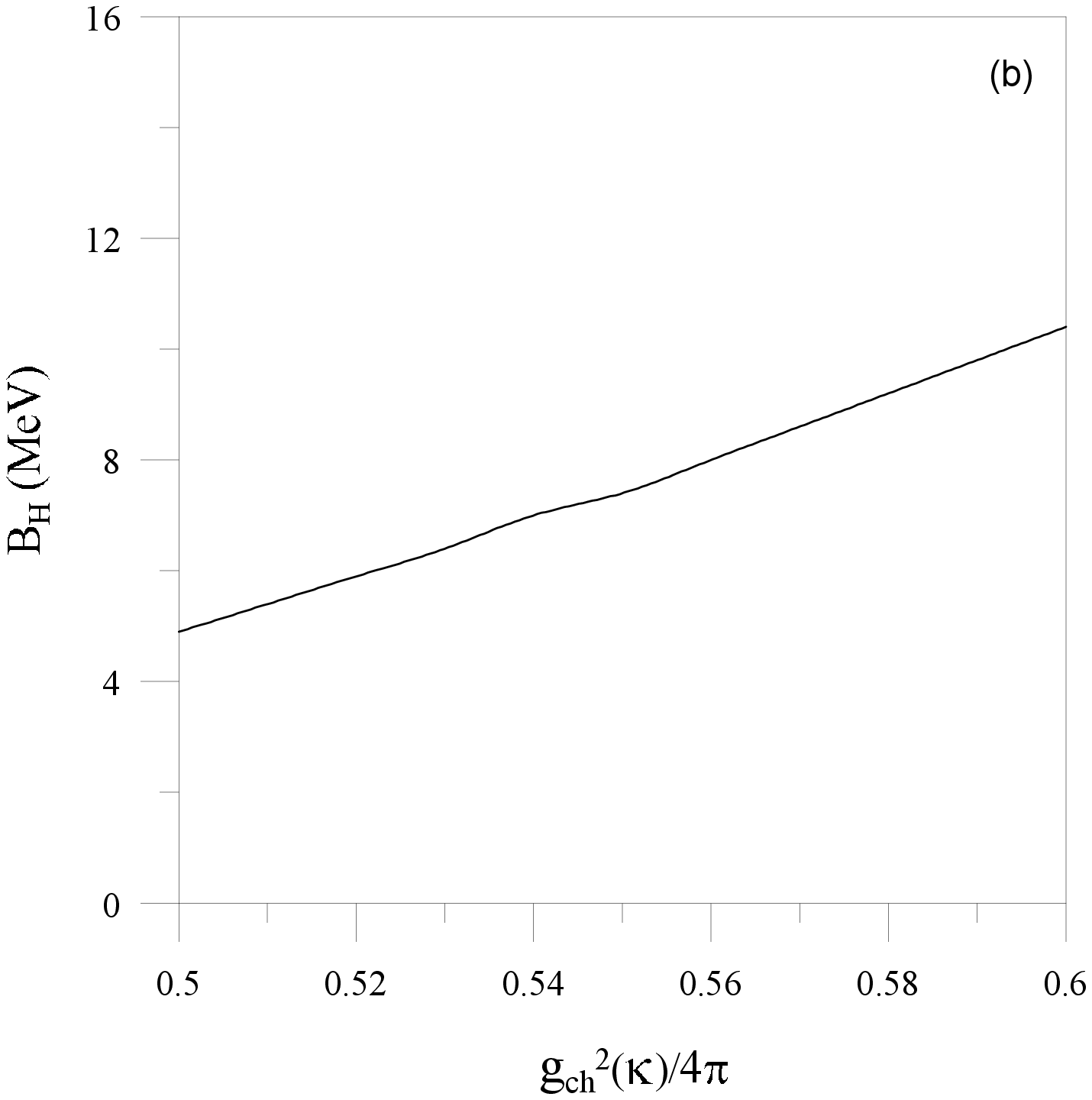,height=4.5in,width=3.1in}}
\vspace*{-3.8cm}
\caption{Dependence of $B_H$ on the quark--meson exchange couplings. See text for details. \label{DepG}}
\end{center}
\end{figure}

In our model, the $\sigma_0$ is very attractive and provides by itself the largest values for $B_H$, from 5.4 MeV to 8.3 MeV. Let us notice that in a pure $\sigma_0$ model the two transition potentials to $N \Xi$ would not be possible, since they are due to the scalar meson exchanges. Then the $\Lambda\Lambda$ would not be coupled to $N \Xi$ and so the two--channel calculation would give the same result as the pure $\Lambda\Lambda$, as can be checked in Table~\ref{contrib}. When adding the $\sigma_0$ to the OGE + PSE terms, an unbound but near threshold H dibaryon is obtained in all cases. As in Ref.~\cite{Shen:1999pf}, quite close to our treatment, the $\sigma_0$ is the only attractive piece. The contribution coming from the exchange of the scalar octet is weakly attractive or repulsive, as can be seen in Table~\ref{contrib}. However, when supplementing the OGE + PSE + $\sigma_0$ potential with the octet of scalars, a bound state appears in the two--channel calculation with $B =$ 1.6 MeV  and $B=$ 7.0 MeV in the three--channel case. It is exactly at this point where the important role of the scalars in our model lies, or more precisely, that of the $\kappa$ exchange. This exchange piece manages to enlarge the binding energy without being attractive itself because it is the main piece in the $\Lambda\Lambda$--$N\Xi$ and $N\Xi$--$\Sigma\Sigma$ transitions. It redistributes part of the flux to the much more attractive $N\Xi$ and $\Sigma\Sigma$ channels, originating thus an appreciable increase in the binding. Similar results have been found in the quark model framework
of Ref.~\cite{Fuj07}.

\subsection{Parameter dependence}
\noindent
As deduced from our discussion above, the binding energy of the dibaryon depends mostly on the $\sigma_0$ and $\kappa$  exchange potentials. Therefore it is interesting to evaluate the dependence of the binding on the parameters involved in these two interactions, more precisely on their masses and coupling constants. Varying the $\sigma_0$ and particularly the $\kappa$ coupling constants along a reasonable range changes the binding energy as seen in Fig.~\ref{DepG}, being these changes larger in the first case. In both cases $B_H$ grows with the coupling constant. Indeed it can be checked that the change on $\rm g_{\rm ch}$, that affects the exchange of every meson, produces quite the same results as the change on $\rm g (\sigma_0)$ exclusively.

\begin{figure}[t]
\begin{center}
\vspace*{-1.3cm}
\hspace*{-0.4cm} \mbox{\psfig{figure=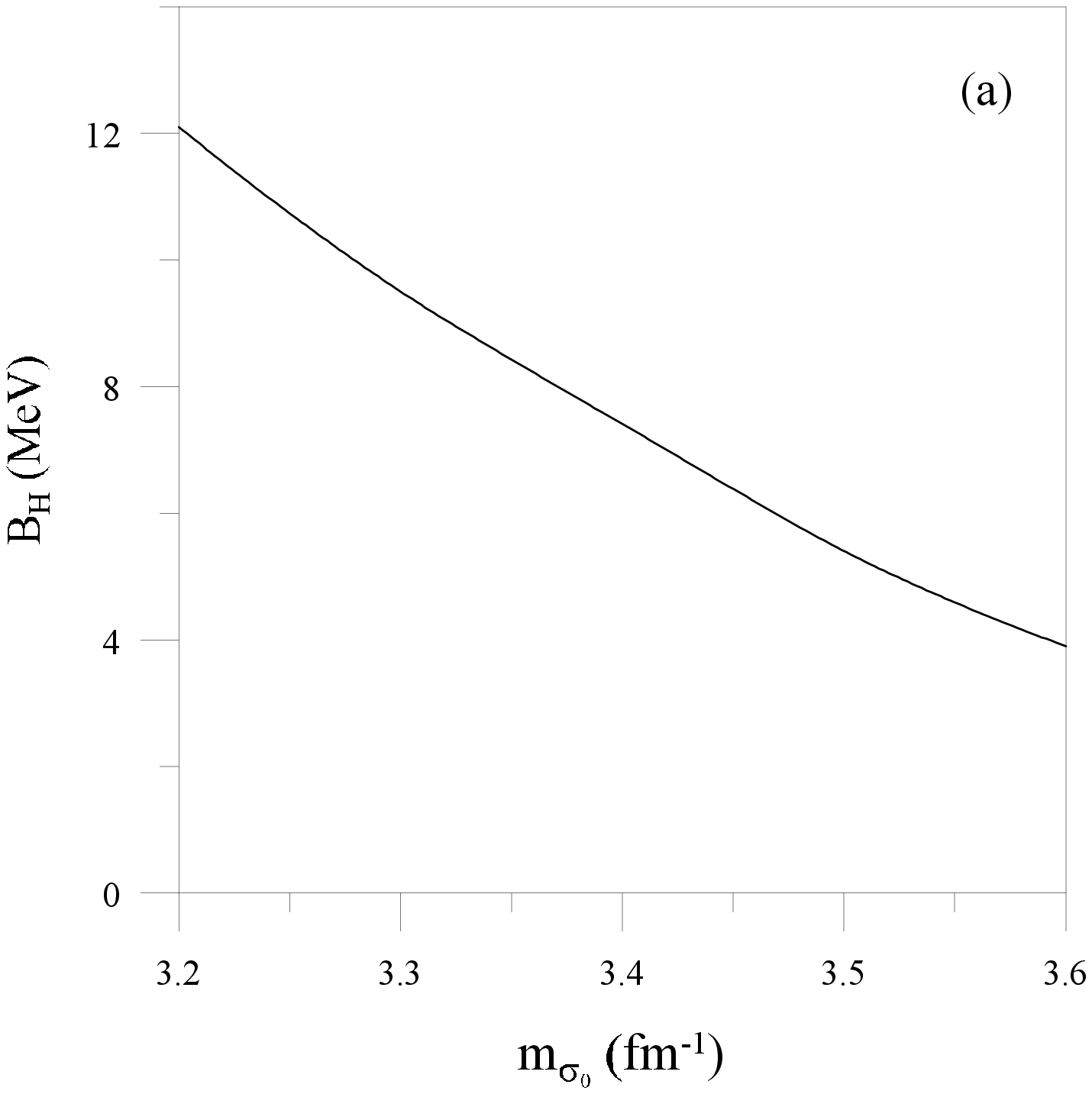,height=4.5in,width=3.1in}}
\hspace*{0.4cm}
 \mbox{\psfig{figure=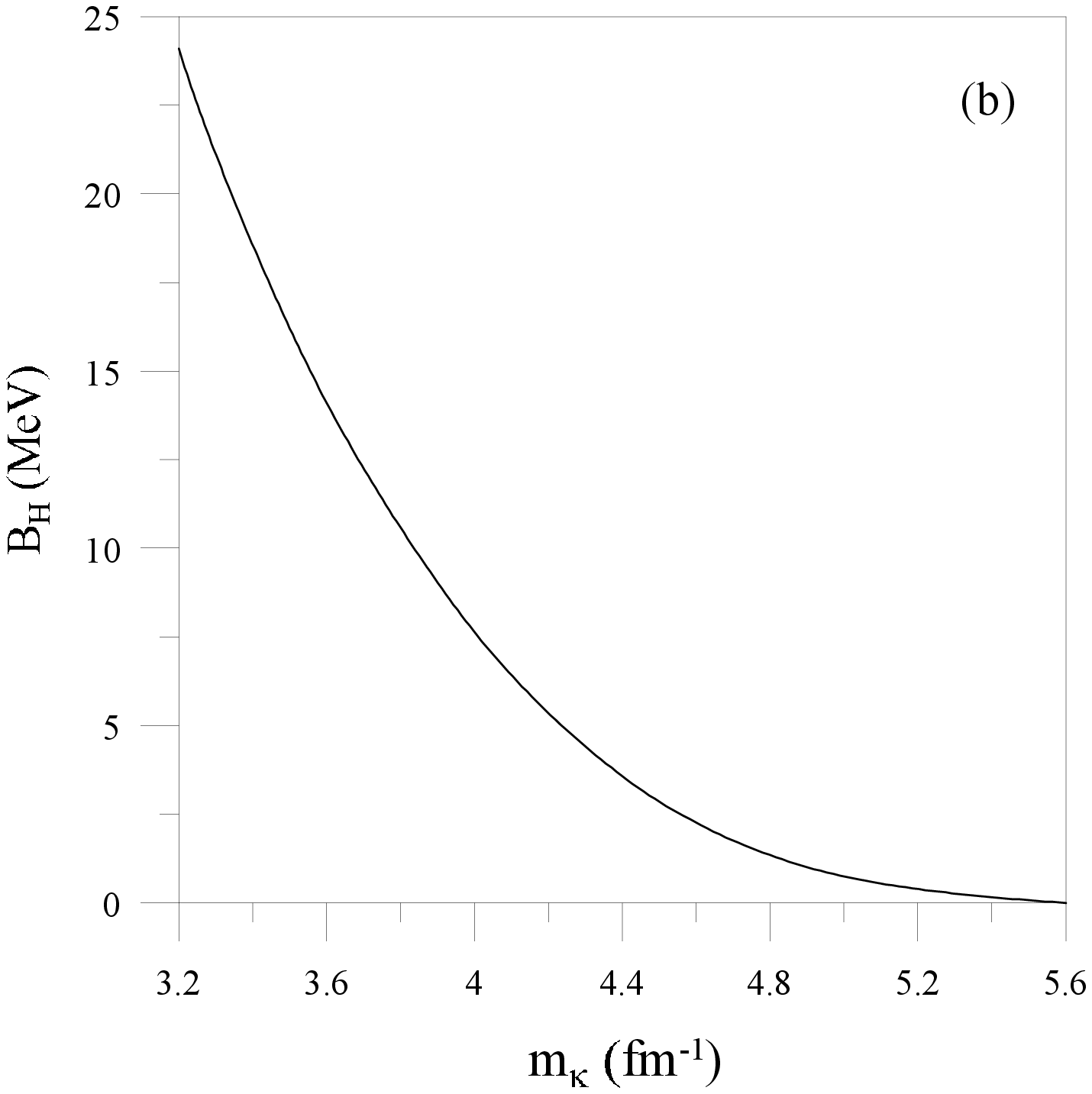,height=4.5in,width=3.1in}}
\vspace*{-3.8cm}
\caption{Dependence of $B_H$ on  the masses of (a) the $\sigma_0$ and (b) $\kappa$ mesons. \label{DepM}}
\end{center}
\end{figure}

The dependence on the masses is opposite, see Fig.~\ref{DepM} for details. The binding decreases as the masses get larger, due to the shorter range of the attraction. This behavior is more pronounced in the case of $m_{\sigma_0}$. $B_H$ ranges approximately from 4 MeV to 12 MeV for $m_{\sigma_0} \in (3.2 , 3.6) $fm$^{-1}$. Both dependence features (masses and couplings) are general, and similar qualitative results were also found in~\cite{Shimizu:1999fq}.

\subsection{Flavor symmetry breaking}
\noindent
The flavor symmetry breaking (FSB) is expected to be an important effect in the H dibaryon due to the presence of two strange quarks. The present study has been made under the assumption that SU(3) is broken. Several sources of symmetry breaking are  taking part in our work. First of all, the strange quark has been assigned a mass different to the light quark masses ($m_s$ = 555 MeV $\neq m_{u,d} = $ 313 MeV). It enters in the expressions of the OGE and pseudoscalar potentials through the factor $1/m_i m_j$. 
However, the differentiation of the $s$--quark in the wave function through the harmonic oscillator parameter $b_s$ gives the larger source of FSB. Besides giving different values for the orbital matrix elements, it implies the suppression of a number of diagrams because light and strange quarks, as distinguishable particles, cannot be exchanged.

\begin{figure}[t]
\begin{center}
\vspace*{-1.3cm}
\hspace*{-0.4cm} \mbox{\psfig{figure=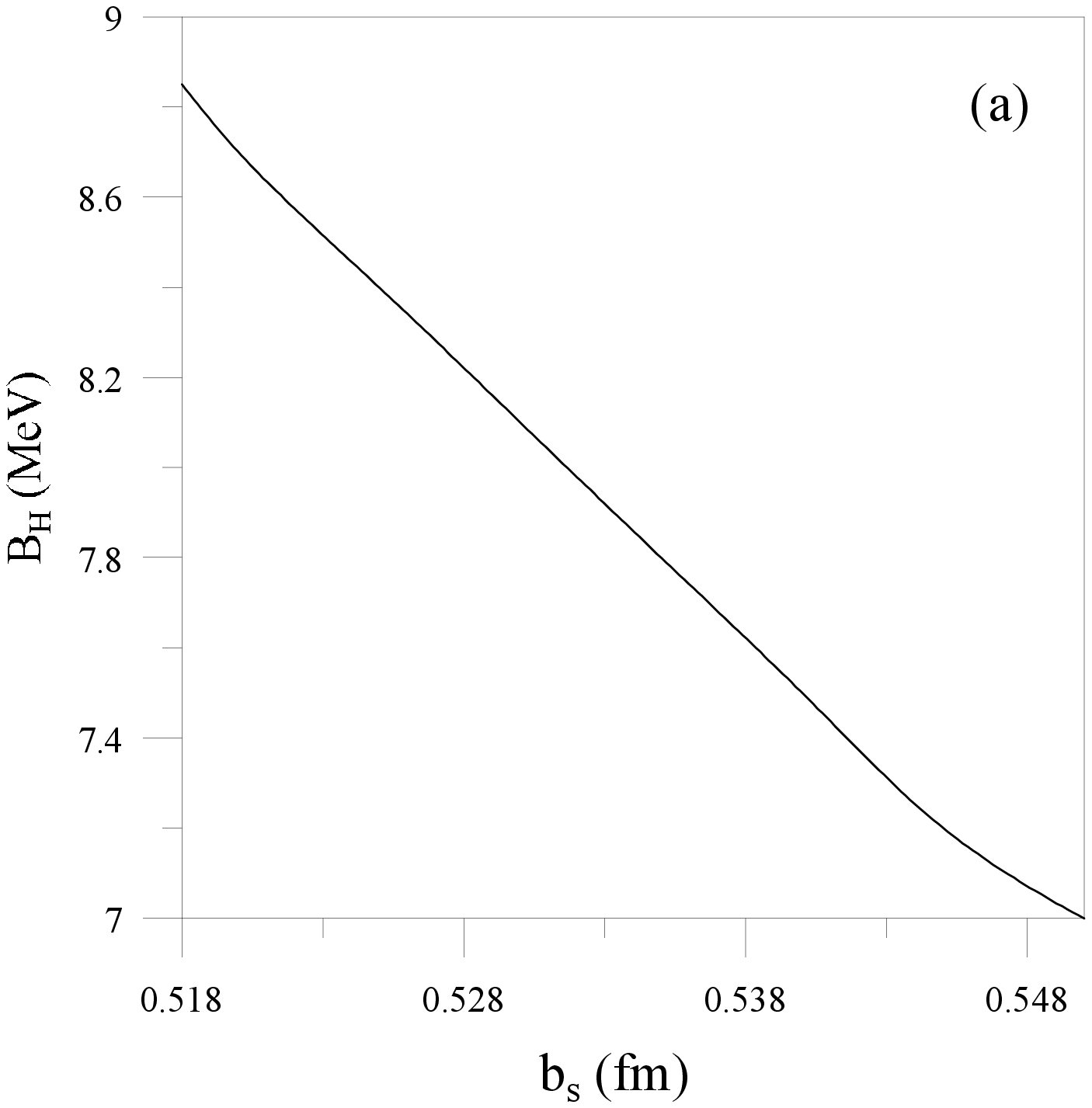,height=4.5in,width=3.1in}}
\hspace*{0.4cm}
 \mbox{\psfig{figure=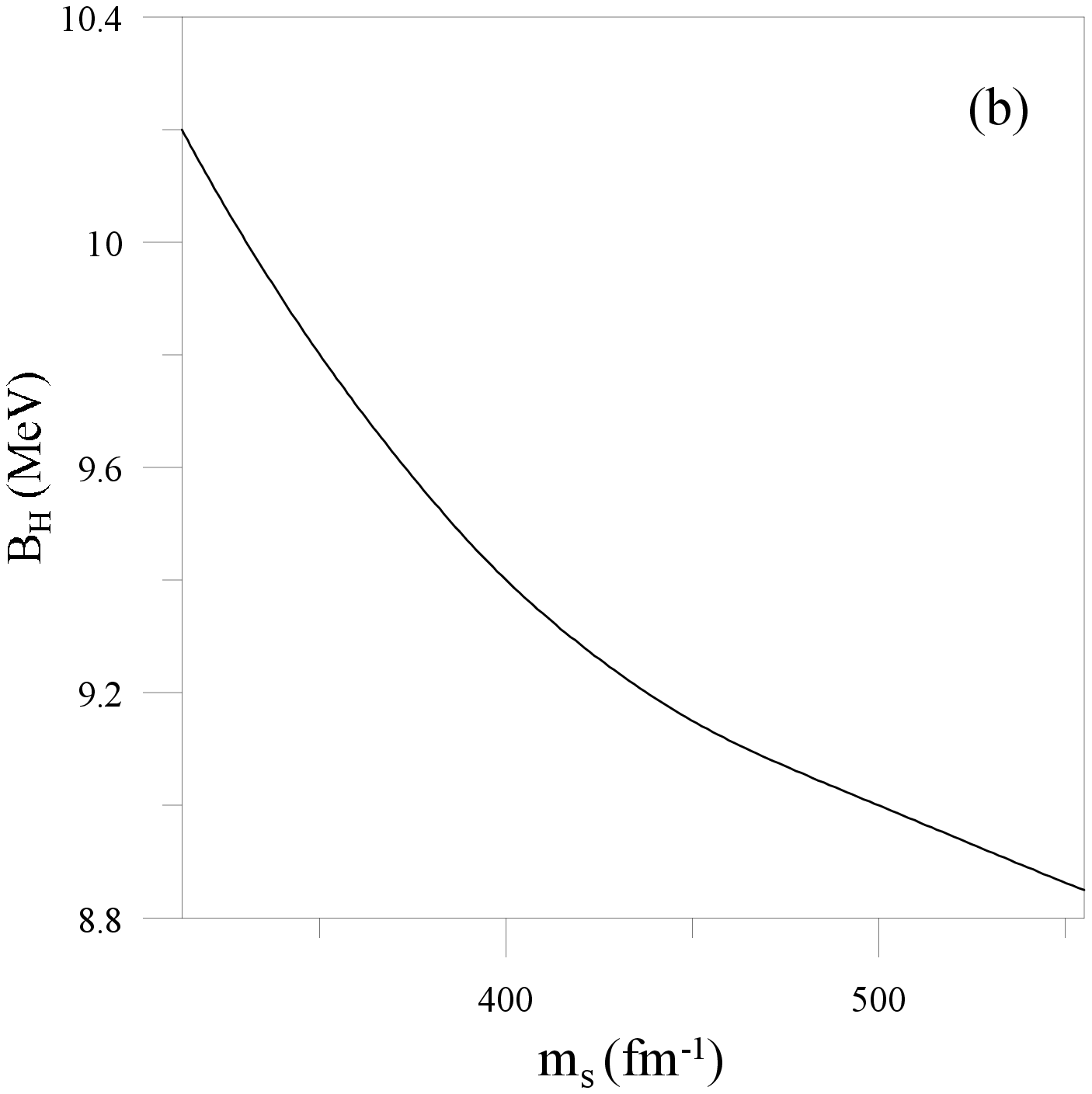,height=4.5in,width=3.1in}}
\vspace*{-3.8cm}
\caption{Dependence of $B_{H}$ on the strange quark (a) harmonic oscillator parameter and (b) mass. \label{par}}
\end{center}
\end{figure}

In order to roughly evaluate the effect of FSB on the H dibaryon, the binding energy has been recomputed when some sources of symmetry breaking have been eliminated, i.e., for $m_s = m_{u,d} = $ 313 MeV and $b_s = b$. We have variated in Fig.~\ref{par} the harmonic oscillator parameter as well as the mass of the strange quark, between their values for exact SU(3) and their true values in our model. One can see how the binding energy increases as one approaches the exact SU(3) limit. As a consequence, the binding is increased by around 45$\%$, giving $B_H =$ 10.2 MeV. Thus, we can conclude that FSB lowers the attraction. A similar conclusion was obtained in Ref.~\cite{Oka:1983ku}, through a RGM calculation with only OGE and confinement potentials. There, the binding energy drastically changed from 38 MeV for exact SU(3) to 26 MeV above threshold when the FSB effect was incorporated. A similar qualitative result was found by Ref.~\cite{Shimizu:1999fq}, where the authors concluded that the symmetry breaking diminished the strong attraction coming from the chromomagnetic interaction. From chiral effective field theory~\cite{Haidenbauer:2011ah}, also the bound state found for exact SU(3) disappears when using physical values for the baryon and meson masses.

A closer inspection of every component of our potential in all the six channels contributing to H allows us to conclude that the effect of FSB is to diminish the strength of the potential, be it repulsive or attractive. Therefore, depending on the combination of the contributions in any channel, the total potential will be larger or smaller than the one with FSB, but no other general conclusion can be inferred. In our particular case, it turns out that $\Lambda\Lambda$ and $\Sigma\Sigma$ hardly change, whereas $N\Xi$  gains attraction when approaching SU(3). As for the transitions, the only appreciable change takes place in the $\Sigma\Sigma$--$N\Xi$ interaction, that acquires twice as much attraction for the exact SU(3) case. This result is in contrast with the one found in~\cite{Nakamoto:1997gh}, where the $\Lambda\Lambda$--$N\Xi$ interaction is strongly dependent on a FSB parameter.

\subsection{Spatial configuration}
\noindent
At first sight there are two types of possible configurations for the H dibaryon: the six--quark cluster and the two--baryon cluster. As we have already mentioned the first one was employed in the original work~\cite{Jaffe:1976yi}, getting a bound H. However latter calculations found that a dibaryon with only the $(0s)^6$ configuration would not bind~\cite{Shen:1999pf}. The spatial distribution was also investigated in Ref.~\cite{Shimizu:1999fq}, where an extended resonating group method was employed in order to account for the possibility of a change in the baryon wave functions. Again, the dibaryon did not bind in the $(0s)^6$ configuration, being thus necessary to consider less compact configurations, due to the medium range attraction. When enlarging the size of the wave function, the dibaryon became more attractive, being $B_H=$ 18.2 MeV for the stable solution. The probabilities however did not drastically change from the more compact ($(0s)^6$) to the more spread (two--cluster) configuration. In the latter case they were found to be $P_{\Lambda\Lambda} = $ 0.216, $P_{N \Xi}=$ 0.543 and $P_{\Sigma\Sigma}=$ 0.242. This result is slightly different from that in the flavor singlet state, Eq.~(\ref{hbcon}). The small difference justifies the perturbative treatment employed, since the wave function was first postulated to be the flavor singlet when performing the RGM calculation. 

Usually, the flavor singlet wave function, Eq.~(\ref{hbcon}), is first postulated and a perturbative variational calculation is afterwards performed when making use of the RGM treatment. This is not the case of our work, since the coefficients of the baryon--baryon components of the flavor wave function are obtained as an output of the calculation, without making further assumptions. The probabilities we got are $P_{\Lambda\Lambda} = $ 0.177, $P_{N\Xi}=$ 0.446 and $P_{\Sigma\Sigma}=$ 0.377. They are quantitatively similar to those of the flavor singlet, from what we can infer that in our model the baryon--baryon wave function is at first approximation SU(3) symmetric, being the difference between both due to the flavor symmetry breaking effects.
\begin{figure}[tbp]
\begin{center}
\vspace*{-2.3cm}
\hspace*{-0.4cm} \mbox{\psfig{figure=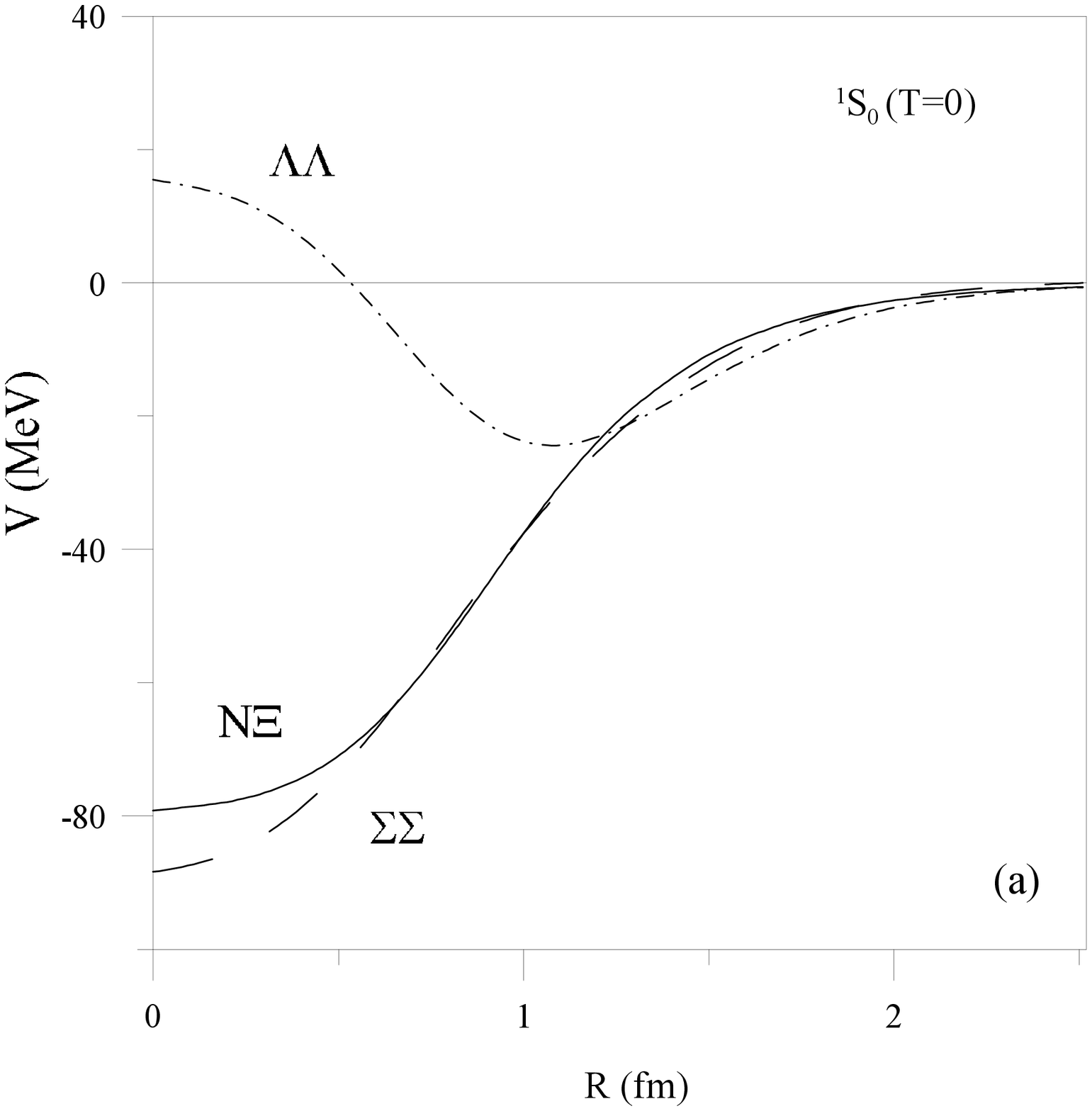,height=3.5in,width=2.5in}}
\hspace*{0.4cm}
 \mbox{\psfig{figure=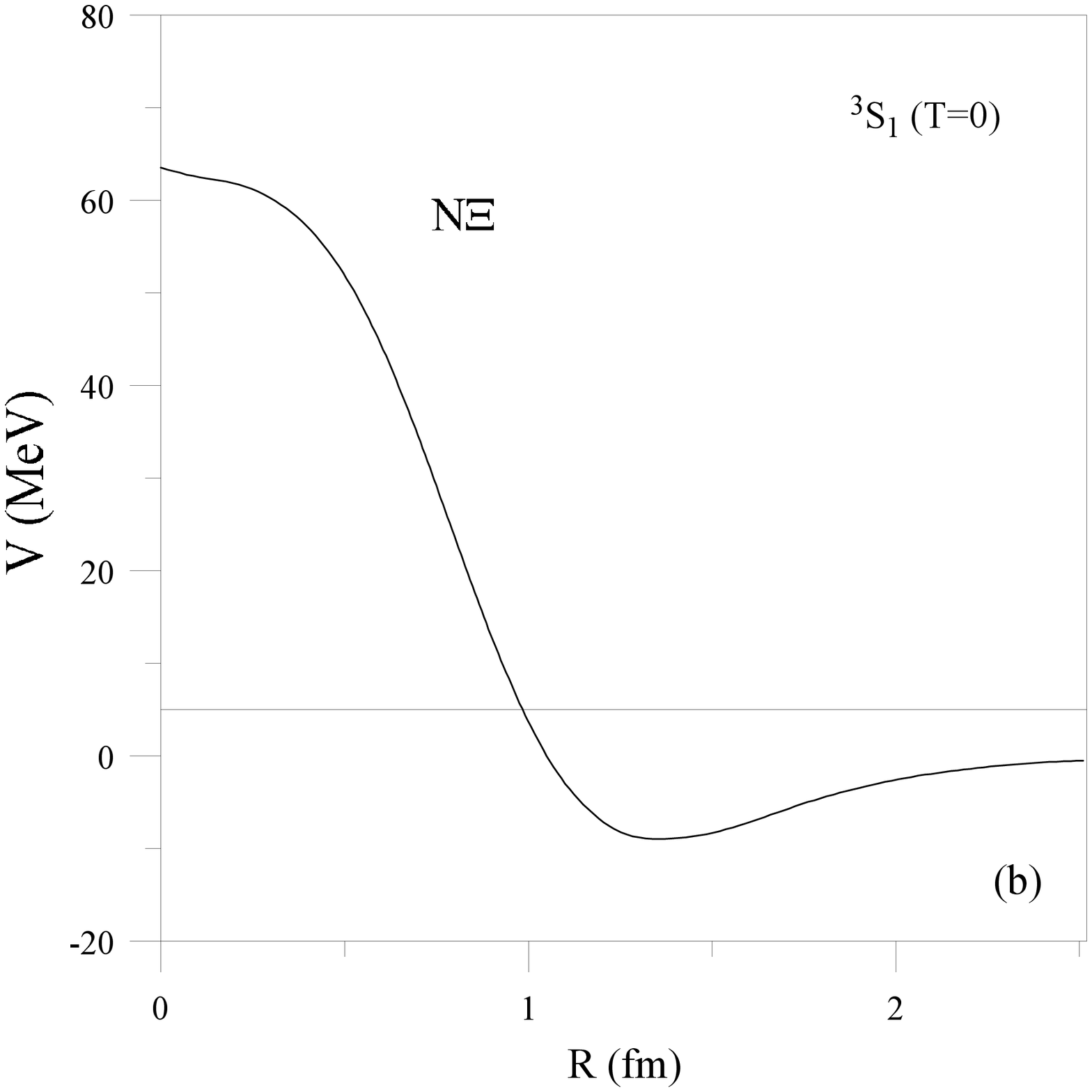,height=3.5in,width=2.5in}}
\vspace*{-1.3cm}
\caption{Interacting potentials contributing to the $T = 0$ channels. (a) $S = 0$ and (b) $S = 1$. \label{t0}}
\end{center}
\end{figure}
\subsection{Other states in $\hat{S} = -2$}
\noindent
In order to properly understand the $\hat{S} = -2$ sector, the analysis performed on the H dibaryon and the results obtained should be complemented studying the other spin--isospin channels. From such a complete analysis, more general and powerful conclusions may arise.

Unlike the $^1S_0 (T = 0)$ H dibaryon channel, the $^3S_1 (T = 0)$ is found to be very weakly attractive. Only the $N \Xi$ interaction is present. The $\Lambda\Sigma$ system does not couple to $T = 0$  and the wave function of $\Lambda\Lambda$ and $\Sigma\Sigma$ is antisymmetric, and therefore vanishes, for $L$ even. 
Since $\Lambda\Lambda$ is forbidden, there are no lighter states and therefore the $N \Xi$ state could decay weakly. The $^3S_1 (T = 0) N\Xi$ potential, Fig.~\ref{t0}($\rm b$), exhibits a repulsive core, due to the strong OGE interaction that compensates by far the attraction coming from the $\sigma_0$. At intermedium range the attractive feature survives but is not enough to form a bound state. The weak attraction in $N \Xi$ can be enhanced if the tensor force is turned on, however a bound state is still not possible. This qualitative behavior coincides with that of Refs.~\cite{Nakamoto:1997gh,Koike:1989ak} and Nijmegen models D and F \cite{Nagels:1976xq}. The big difference from the H channel is an example of the importance of the flavor dependence interaction that Model II incorporates.  A summary of the
predictions for the scattering lengths of the $\hat{S}=-2$ two-baryon systems with modern versions of the Nijmegen
potential~\cite{Nij06}, the J\"ulich potential~\cite{Pol07}, and the work by Fujiwara {\it et al.}~\cite{Fuj07}
can be found in Table 2 of Ref.~\cite{Valcarce:2010kz}. 
%
\begin{figure}[tbp]
\begin{center}
\vspace*{-2.8cm}
\hspace*{-0.4cm}\mbox{\psfig{figure=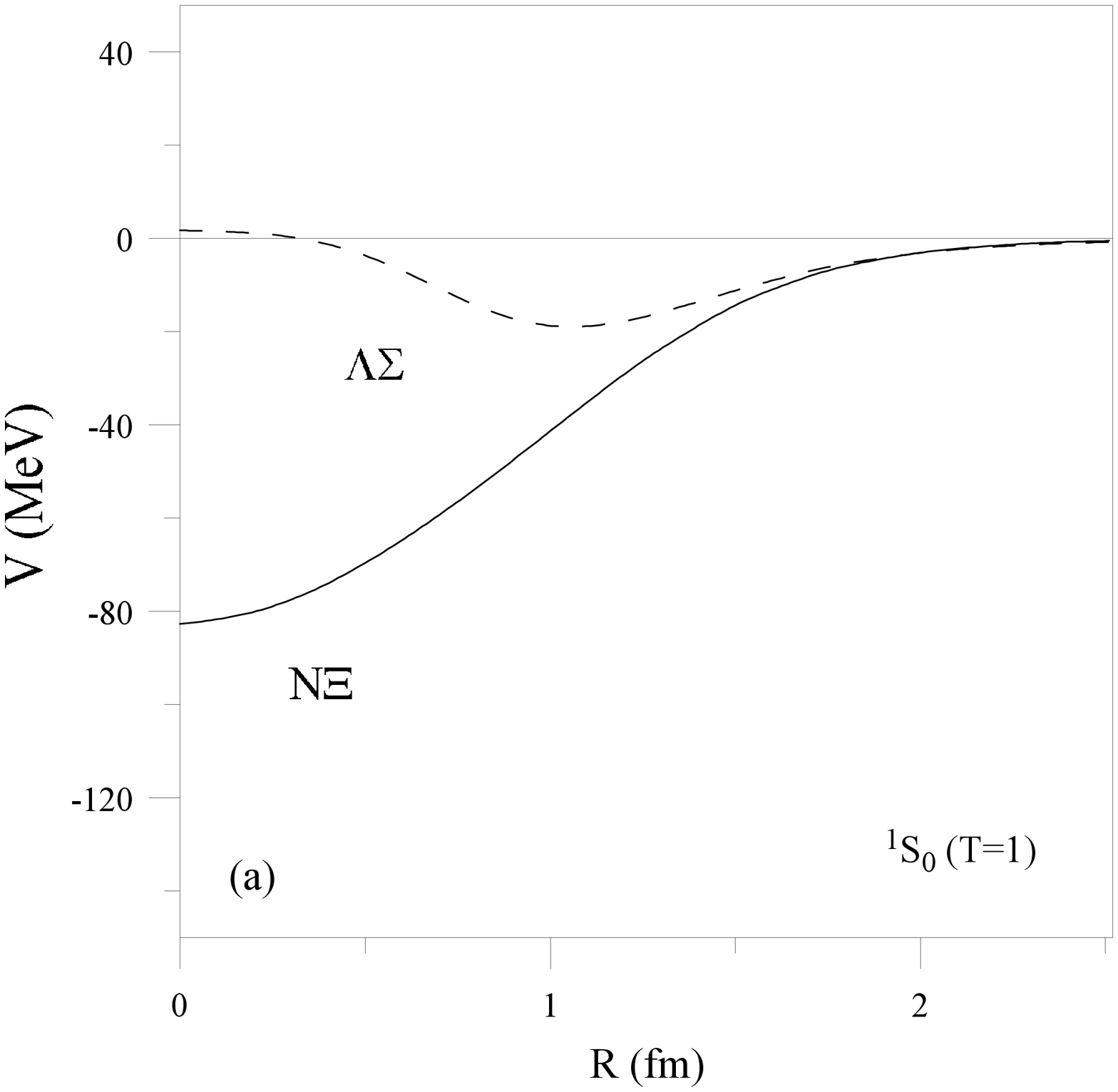,height=3.5in,width=2.5in}}
\hspace*{0.4cm}
\mbox{\psfig{figure=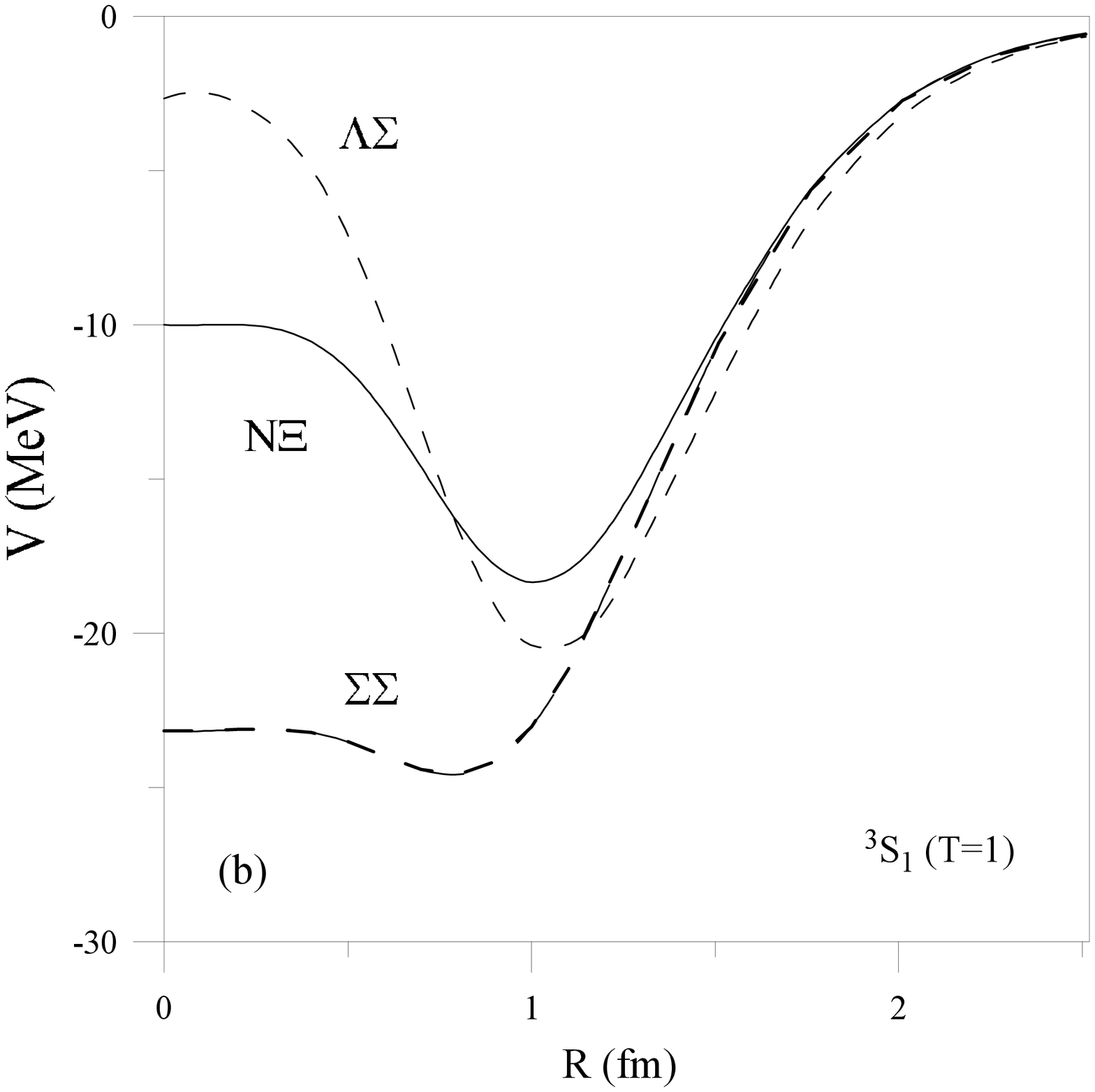,height=3.5in,width=2.5in}}
\vspace*{-0.9cm}
\caption{Interacting potentials contributing to the $T = 1$ channels. (a) $S = 0$ and (b) $S = 1$. \label{t1}}
\end{center}
\end{figure}
%

The potentials in the two $T = 1$ channels are all attractive. Regarding the $^1S_0$, both $N \Xi$ and $\Lambda\Sigma$ interactions are strongly attractive, as can be seen in Fig.~\ref{t1}($\rm a$).  
Such a strong attraction provides for the coupled system a binding energy of 4.8 MeV. However this is not the case in Ref.~\cite{Nakamoto:1997gh}, where all the potentials with $T = 1$ are repulsive. As for the $^3S_1$ channel, there are three coupled interactions: $N \Xi$, $\Lambda\Sigma$ and $\Sigma\Sigma$. The three of them are moderately attractive and look very similar, as can be seen in Fig.~\ref{t1}($\rm b$). The Nijmegen models D and F~\cite{Nij06} also predict attractive interactions. However, no bound state appears in our model.

\section{Summary}
\label{secIV}
We have performed the first calculation of the H dibaryon in a model constrained by the elastic and inelastic $\Lambda N, \Sigma N$, $\Xi N$ and $\Lambda\Lambda$ cross sections. Special interest has been devoted to analyze the contribution of the different pieces of the interaction and to the effect that the addition of channels to the $\Lambda\Lambda$ system produces on the binding. We obtained a bound H dibaryon, with $B_H =$ 7 MeV, compatible with a plausible extrapolation of recent lattice QCD results and with the Nagara event, the most stringent restriction known so far. The scalar octet exchange, although not giving an attractive contribution by itself, plays a key role as it is the main ingredient of the transition potentials that connect $\Lambda\Lambda$ to the more attractive $N \Xi$ and $\Sigma\Sigma$ states. The probabilities $P_{\Lambda\Lambda} = $ 0.177, $P_{N\Xi}=$ 0.446 and $P_{\Sigma\Sigma}=$ 0.377 are quite similar to those of the flavor singlet so that our wave function is at first approximation SU(3) symmetric. 
Finally, a bound state has also been found in the $\hat{S}=-2$, $(T,S)=(1,0)$ channel, with a binding energy of 4.8 MeV, thus smaller than $B_H$. 
The abundance of events foreseen in a near future and/or the improvement on the lattice calculations will help us to advance in our knowledge of the mechanisms that play important roles in the dynamics of the H dibaryon. For such a task, a detailed theoretical study as the one presented here could be relevant.

\acknowledgments This work has been partially funded by Ministerio de
Educaci\'on y Ciencia and EU FEDER under Contract No. FPA2010--21750--C02--02 and by the
Spanish Consolider--Ingenio 2010 Program CPAN (CSD2007--00042).


\end{document}